\documentclass[12pt]{iopart}

\usepackage[svgnames]{xcolor}      
\usepackage{times}
\usepackage{geometry}
\usepackage[utf8]{inputenc}
\usepackage{enumitem,amssymb}
\usepackage{ragged2e}
\usepackage{array} 
\usepackage{booktabs} 
\usepackage{graphicx}
\usepackage{comment}
\usepackage{multicol}
\usepackage[caption=false]{subfig}             
\definecolor{xlinkcolor}{cmyk}{1,1,0,0}
\usepackage{url}
\usepackage{booktabs}
\usepackage{bm}
\usepackage[normalem]{ulem}
\PassOptionsToPackage{table,dvipsnames}{xcolor}

\usepackage{titlesec}              
\usepackage{lmodern}  
\usepackage[numbers,square,sort&compress]{natbib}
\usepackage[
colorlinks=true,    
linkcolor=xlinkcolor,     
citecolor=xlinkcolor,     
filecolor=xlinkcolor,  
urlcolor=xlinkcolor,      
final=true
]{hyperref}

\emergencystretch=\maxdimen
\usepackage{enumitem}
\setenumerate{itemsep=0mm}

\begin{document}

\title[Binary Neutron Stars from the Moon]{Binary Neutron Stars from the Moon: Early Warnings and Precision Science for the Artemis Era}

\author{Anjali B. Yelikar$^1$ and Karan Jani$^1$}
\address{$^1$Vanderbilt Lunar Labs Initiative, Vanderbilt University, 6301 Stevenson Center Lane, Nashville, TN 37235, USA}
\ead{anjali.yelikar@vanderbilt.edu}

\begin{abstract}
Binary neutron star mergers are unique probes of matter at extreme density and standard candles of cosmic expansion. The only such event observed in both gravitational waves and electromagnetic radiation, GW170817, revealed the origin of heavy elements, constrained the neutron star equation of state, and provided an independent measurement of the Hubble constant. Current detectors such as LIGO, Virgo, and KAGRA capture only the final minutes of inspiral, offering limited advance warning and coarse sky localization. In this study, we present a comprehensive analysis of binary neutron star signals for lunar-based gravitational-wave observatories (LILA, LGWA, GLOC) envisioned within NASA's Artemis and Commercial Lunar Payload Services programs, and compare their performance with current and next-generation Earth-based facilities. For GW170817-like sources, we find that lunar detectors can forecast mergers weeks to months in advance and localize them to areas as small as 0.01 deg$^{2}$, far beyond the reach of terrestrial detectors. We further show that lunar observatories would detect on the order of 100 well-localized mergers annually, enabling coordinated multi-messenger follow-up. When combined in a multi-band LIGO+Moon network, sky-localization areas shrink to just a few arcsec$^{2}$, comparable to the field of view of the James Webb Space Telescope at high zoom. Multi-band parameter estimation also delivers dramatic gains: {binary} neutron star mass-ratio uncertainties can be measured with $\sim0.1\%$ precision, spin constraints to 0.001$\%$  with luminosity distance errors to 1$\%$ level, enabling precision measurements of the equation of state and the cosmic expansion rate. Our results demonstrate that lunar gravitational-wave observatories would revolutionize multi-messenger astrophysics with binary neutron stars and open a unique discovery landscape in the Artemis era. 
\end{abstract}

\section{Introduction} \label{sec:intro}

Binary neutron star (BNS) systems have now been observed in both the electromagnetic (EM) and gravitational wave (GW) windows of our Universe. The Hulse-Taylor binary pulsar~\citep{1975ApJ...195L..51H} provided the first EM observations of neutron stars in a binary system, along with indirect evidence of the existence of GWs. The first direct observation of GWs from a BNS system was the event GW170817~\citep{LIGOScientific:2017vwq,LIGOScientific:2018hze}, which also had EM follow-up observations, which provided proof that such systems are progenitors of short gamma-ray bursts (GRBs)~\citep{LIGOScientific:2017ync}. Since then, the LIGO-Virgo-KAGRA (LVK) GW detectors also observed GW190425~\citep{LIGOScientific:2020aai,Raaijmakers:2021slr}, but no counterpart EM signals were found, and the masses of the neutron stars in the binary were also found to be heavier than the known Galactic BNS population~\citep{Farrow:2019xnc}.

\begin{figure}[t!]
    \centering
    \includegraphics[width=1\linewidth]{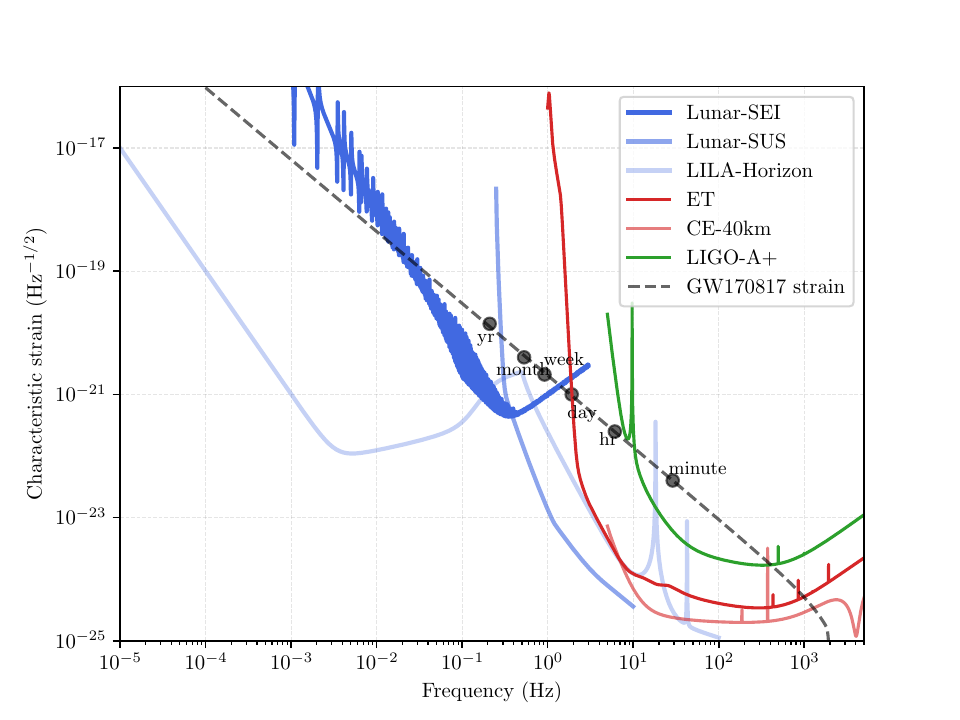}
    \caption{Sensitivity curves of various detectors, along with GW strain of a GW170817-like signal with annotated times to merger.}
    \label{fig:sensitivity}
\end{figure}

The fourth observing run of LVK began in May 2023; however, only two BNS mergers have been confirmed from GW Transient Catalog(GWTC)-1 through GWTC-4. The current GW detectors are also sensitive only to the inspiral regime of coalescing neutron stars and cannot observe the post-merger of such collisions; the remnant of GW170817 still eludes us~\citep{LIGOScientific:2017fdd}. Observing BNS in the GW domain can help in understanding the equation of state of matter~\citep{Raaijmakers:2019dks,Kedia:2024hvw} and the structure of a NS~\citep{LIGOScientific:2018cki,Zhu:2025dea} better when combined with information from EM and nuclear physics models. BNS also serve as a cosmic distance ladder, and help constrain the Hubble constant~\citep{Nissanke:2013fka}, as demonstrated by the follow-up work from GW170817/AT2017gfo observation~\citep{LIGOScientific:2017adf,Palmese:2023beh}. An independent measurement of the Hubble constant is particularly relevant given current tension between local and early-Universe estimates~\citep{Feeney:2018mkj,Freedman:2021ahq,Freedman:2025nfr,Simpson:2025kfn}. 

We can ensure a joint GW+EM detection of coalescing neutron stars if one has enough early warning about the accurate localization of the source so as to alert the EM telescopes. Currently in the LIGO-Virgo-KAGRA detector bands, this early warning is just the order of a few minutes and the sky-localization area depends on how many of the detectors were in observing mode at the time of detection. One can obtain better early warning alerts if the sensitivity of the GW detectors in the low-frequency region could be increased; however, this is currently limited by the seismic noise. There are proposed third-generation (3G) GW detectors such as Cosmic Explorer (CE)~\citep{2017CQGra..34d4001A,2019BAAS...51g..35R,2021arXiv210909882E} in the US and the Einstein Telescope (ET)~\citep{Punturo_2010, Hild:2010id} in Europe that have better sensitivity in the 5-20 Hz frequency band which would help in obtaining early warning alerts of the order of a $\sim$30 minutes-1 hr. 

Even though the 3G detectors have a projected sensitivity in the 5-20 Hz region, they will still be limited by the seismic noise on Earth. A proposal has been to build GW detectors on the Moon owing to its lower gravity and seismic noise levels. The Moon also has natural craters which can serve as the best sites to place detectors with arms as long as 40km, without having to dig underground as is proposed to be done for 3G detectors on Earth, to correct for Earth's curvature. There are currently {three} GW detectors proposed on the Moon- the Lunar Gravitational Wave Antenna (LGWA)~\citep{LGWA(2021), LGWAWP}, Gravitational-Wave Lunar Observatory for Cosmology (GLOC)~\citep{GlocJani2020arXiv200708550J, Jani_Artemis2022,ballmer2022snowmass2021cosmicfrontierwhite} and Laser Interferometer Lunar Antenna (LILA)~\citep{Jani:2025uaz,Creighton:2025kth,Panning:2025jcx,Shapiro:2025oqa}. LILA is being proposed as part of NASA's Artemis Lunar Exploration Program~\citep{Trippe(2024)LPICo}. These Lunar GW detectors have a projected early warning alert of a few days to several months. 

Observing the GW signal from the remnant (or post-merger) of a BNS is not possible with the current GW detectors. The 3G Earth-based detectors and Lunar detectors would have a better high-frequency sensitivity up to a few kHz. Detecting a signal from the remnant~\citep{Faber:2012rw,Bernuzzi:2020tgt} can also help in understanding the equation-of-state of the NS, as well as help understand the boundaries of the various remnant stages, such as supramassive NS, hypermassive NS, or whether BNS systems promptly collapse to a black hole. The understanding of the dense matter equation of state will also improve from more observations of BNS in this high-frequency region.  
Overall, from a GW instrumentation perspective, to further the BNS science, the scientific community requires improvements in the following areas for:
\begin{itemize}
    \item \textbf{Multimessenger} studies: Need better low-frequency sensitivity for early alerts and better sky location, inclination and distance measurement.
    \item \textbf{Equation-of-state} studies: Need better high-frequency sensitivity ($\sim$kHz) along with low-frequency improvements for better mass, spin, and tides constraints. 
\end{itemize}

In this study, we discuss the science case of BNS detection for the Lunar GW detectors and compare it against current/proposed 3G detectors on Earth. Section \ref{sec:methods} describes the various GW detectors and configurations considered, along with parameter estimation techniques. In Section \ref{sec:results}, we report constraints of a GW170817-like signal and also discuss the improvements in constraints when a multi-detector study is conducted, which we refer to as multi-band PE. We also discuss detection rates for a population of BNS mergers. In Section \ref{sec:discussion}, we discuss the implications of how Lunar detectors will help the case of multiband and multimessenger astronomy.

\begin{figure}
    \centering
   \includegraphics[width=0.9\linewidth]{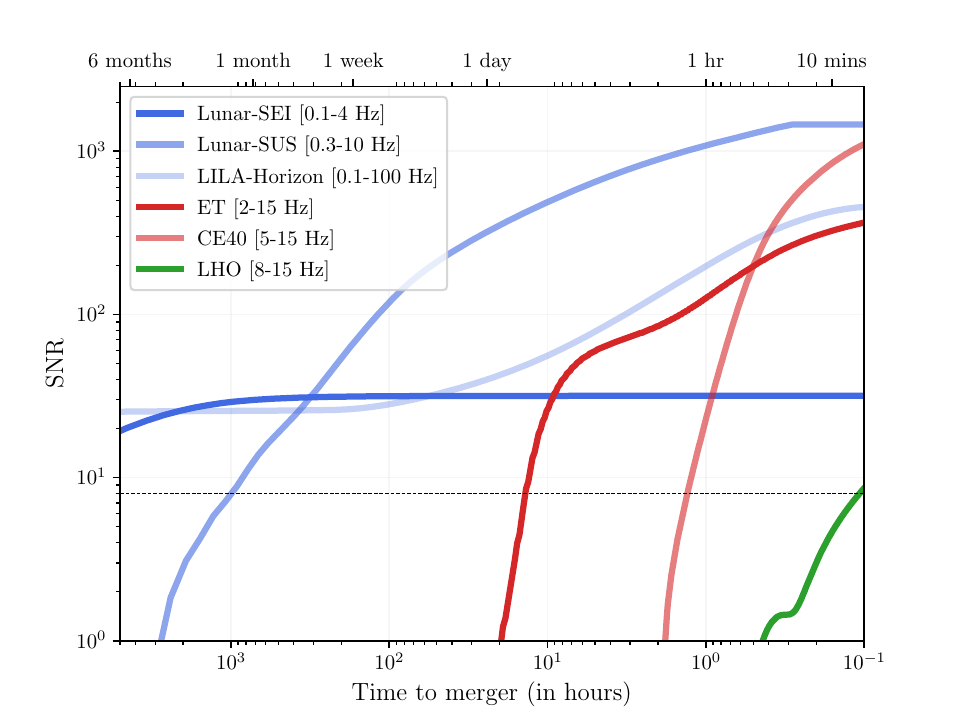}
    \caption{SNR evolution to merger as a function of time (in hours). The black dashed line at SNR=8 shows the threshold that is generally used to claim a GW detection. }
    \label{fig:EW}
\end{figure}

\section{Methodology} \label{sec:methods}

\subsection{Detectors}

In this work, we consider current Earth-based (2G) and proposed Earth (3G) and Moon-based GW detectors. Below, we detail their configuration and frequency sensitivity bands.

\begin{itemize}

\item \textbf{Second-generation (2G)} network of Advanced LIGO~\citep{LIGOScientific:2014pky} detectors at the following sites: Hanford (\texttt{H} or \texttt{LHO}), Livingston (\texttt{L} or \texttt{LLO}) in the USA and Aundha(\texttt{A} or \texttt{LIN}) in India~\citep{LIGO-India} are considered at the A+ sensitivity. They are giant L-shaped Michelson interferometers with 4 km-long arms containing suspended test masses, and are sensitive to GW signals in the frequency range of $\sim$8-2048 Hz. 

\item In the \textbf{third generation (3G)} of Earth-based detectors, we consider the proposed detectors Cosmic Explorer (CE) and Einstein Telescope (ET). We assume CE to be hypothetically located at the coordinates of LIGO-Hanford in the USA with 40km-long and/or 20km-long arms and consider it sensitive in the $\sim$5-2048 Hz frequency band. This will be referred to as \texttt{CE40} or \texttt{CE20} depending on the arm length. In certain cases, we also consider a 20km CE in Australia located in New South Wales, and we denote it as \texttt{CE20(Aus)}. ET is assumed to be located in Europe. It features a triangular design, with each set of arms functioning as a single interferometer, resulting in three effective interferometers in the same configuration. The individual arm lengths are proposed to be 10km, with the detector sensitivity being between $\sim$2-2048 Hz. This will be referred to as \texttt{ET}. 

\item Lunar detectors that use seismometers to monitor the normal modes of the Moon and their response to GWs, such as the Lunar Gravitational Wave Antenna (LGWA), Laser Interferometer Lunar Antenna(LILA)-Pioneer~\citep{Jani:2025uaz,Creighton:2025kth} {and \texttt{LILA-Horizon LF} (low-frequency)}. We will refer to them as lunar-seismic \textbf{Lunar-SEI}. Their projected frequency sensitivity lies between 0.001 and 1 Hz. In our study, LGWA is used as a proxy for measuring GWs through lunar normal modes.

\item Lunar detectors that are laser-based interferometers with suspended test masses, which we will refer to as lunar-suspension \textbf{Lunar-SUS}, such as Gravitational-Wave Lunar Observatory for Cosmology (GLOC) and {\texttt{LILA-Horizon HF} (high-frequency)~\citep{Jani:2025uaz,Shapiro:2025oqa}}. They are projected to be sensitive in the 0.25-10Hz (sub-Hz) frequency region, which is unattainable on Earth. In our study we use GLOC as a proxy for Lunar-SUS and although its proposed sensitivity extends to 2kHz, in this work we consider the gain in signal SNR only up to 10Hz. 

\item {We also discuss the \texttt{LILA-Horizon} detector in this study, which is a combination of a setup to monitor the normal modes of the Moon (\texttt{LILA-Horizon LF})~\citep{Creighton:2025kth} along with laser-based interferometers with suspended test masses (\texttt{LILA-Horizon HF})~\citep{Shapiro:2025oqa}.} 
\\

Figure~\ref {fig:sensitivity} shows the sensitivity of various 2G, 3G, and Lunar detectors, along with the GW170817 strain as a function of frequency. The signal is seen in the \texttt{Lunar-SEI} band a few years before merger {which is also the case for \texttt{LILA-Horizon}}, whereas it enters the \texttt{Lunar-SUS} detector band ($f_{low}$=0.3Hz) starting $\sim$200 days (6.5 months) before merger. The signal enters the \texttt{ET} and \texttt{CE40} bands $\sim$1 day to $\sim$1 hour before the merger, respectively. GW170817 enters the \texttt{LIGO} band only a few minutes before merger. 

\end{itemize}


\begin{table}[h!]
\centering
\renewcommand{\arraystretch}{1.2}
\setlength{\tabcolsep}{8pt}
\begin{tabular}{ll}
\toprule
\textbf{Binary Parameter} & \textbf{Value} \\
\midrule
Primary NS mass $m_{1}\,(M_{\odot})$        & 1.510 \\
Secondary NS mass $m_{2}\,(M_{\odot})$      & 1.255 \\
Binary chirp mass $\mathcal{M}\,(M_{\odot})$ & 1.186 \\
Mass ratio $q$                              & 0.831 \\
\midrule
Spin magnitude of primary $a_{1}$           & 0.005 \\
Spin magnitude of secondary $a_{2}$         & 0.003 \\
\midrule
Tidal deformability of primary $\Lambda_{1}$   & 368.178 \\
Tidal deformability of secondary $\Lambda_{2}$ & 586.549 \\
Combined tidal deformability $\tilde{\Lambda}$ & 456.216 \\
\midrule
Luminosity distance $D_{L}$ (Mpc)           & 43.747 \\
Inclination angle $\iota$ (rad)             & 2.545 \\
Right ascension $\alpha$ (rad)              & 3.446 \\
Declination $\delta$ (rad)                  & $-0.408$ \\
\bottomrule
\end{tabular}
\caption{Source parameters of the GW170817-like BNS system used as a fiducial case in this analysis. 
The component masses, mass ratio, and chirp mass are chosen to match the values inferred for GW170817. 
Spin magnitudes are set to small values consistent with the Galactic BNS population. 
Tidal deformabilities $\Lambda_{1}$, $\Lambda_{2}$, and $\tilde{\Lambda}$ follow the same EOS assumptions as adopted in previous studies of GW170817. 
Extrinsic parameters (luminosity distance, inclination, right ascension, and declination) correspond to the observationally inferred values, and serve as the baseline for Fisher-matrix parameter estimation across Earth- and Moon-based detector networks.}\label{tab:GW170817_params}
\end{table}

\subsection{Parameter estimation}

A coalescing BNS in a quasi-circular orbit can be characterized entirely by ten intrinsic
and seven extrinsic parameters.  By intrinsic parameters, we refer to the binary's  (detector-frame) masses $m_i$ and spins $\chi_{i}$, and any quantities
characterizing matter in the system, $\Lambda_i$. The masses are expressed in terms of quantities chirp-mass ({$\mathcal{M}={(m_{1}m_{2})^{3/5}}/{M^{1/5}}$}) and mass ratio ({$q={m_{2}}/{m_{1}}$}) that better characterize the leading-order dependence of the GW phase than the individual masses $m_i$. 
By extrinsic parameters we refer to the seven numbers needed to characterize its spacetime location and orientation: luminosity distance ($d_{L}$), right ascension ($\alpha$), declination ($\delta$), inclination ($\iota$), polarization ($\psi$), coalescence phase ($\phi_{c}$), and time ($t_{c}$).
We will express masses in solar mass units and
 dimensionless spins in terms of Cartesian components $\chi_{i,x},\chi_{i,y}, \chi_{i,z}$, expressed
relative to a frame with $\hat{\mathbf{z}}=\hat{\mathbf{L}}$ and (for simplicity) at the orbital frequency corresponding to the earliest
time of computational interest. We will use \bm{$\lambda$} and \bm{$\theta$} to refer to intrinsic and extrinsic parameters, respectively: $ \bm{\lambda} : (\mathcal{M},q,\chi_{1,x},\chi_{1,y},\chi_{1,z},\chi_{2,x},\chi_{2,y},\chi_{2,z},\Lambda_{1},\Lambda_{2})$ and
$ \bm{\theta} : (d_{L},\alpha,\delta,\iota,\psi,\phi_{c},t_{c}) $.
In this work, we consider only aligned-spin BNS; hence, $\chi_{1,x},\chi_{1,y},\chi_{2,x},\chi_{2,y}$ are assumed to be 0. {For the tidal constraints, instead of the the individual tidal deformability of each NS, we use the combined tidal deformability $\tilde{\Lambda}$ which best captures the leading order effect of tides on the gravitational waveform. It relates the individual tidal deformabilities and the mass-ratio as $\tilde{\Lambda} = \frac{16}{13} \frac{(1+12q)\Lambda_{1}+(12+q)q^{4}\Lambda_{2}}{(1+q)^5}$. }

\subsubsection{GWFish}

We use the Fisher-matrix formalism~\citep{PhysRevD.46.5236, PhysRevD.57.4535} based package  \texttt{GWFish}~\citep{DupletsaHarms2023} to compute the signal-to-noise ratio (SNR), the horizon distance, and measurement uncertainties on the various parameters of a BNS system. For a gravitational waveform $h(f)$, the SNR ($\rho$)  is defined as follows:

\begin{equation}
\rho^{2} = \int_{f_{low}}^{f_{high}}df \frac{h(f)^2}{S_{n}(f)}
\end{equation}

where $S_{n}(f)$ is the noise power spectral density of the detector, and $f_{low}$ and $f_{high}$ are the start and end frequencies considered for signal integration. Similarly, the \textit{horizon distance} is defined as the farthest luminosity distance from which a given source can be detected above an SNR threshold, assuming an optimal source inclination and sky location.

As a first-order estimate of the constraints on various parameters of a BNS system, we also use \texttt{GWFish} to compute errors on source parameter estimation. This method provides a lower bound (called the Cramér-Rao bound) on statistical errors of intrinsic and extrinsic parameters in the high SNR limit~\citep{Vallisneri:2007ev}. The GW signal from the BNS system is modeled using the aligned-spin tidal waveform model \texttt{IMRPhenomD$\_$NRTidalv2}~\citep{Dietrich:2019kaq}. Assuming this waveform model and a set of given source parameter values, the Fisher information matrix $F_{ij}$ is constructed, where each element is an inner product of the partial derivative of $h(f)$ with respect to $i$-th or $j$-th parameter in the source parameters, weighted by the power spectral density of the detector noise ($S_{n}(f)$). The diagonal components of the inverse of $F_{ij}$ are the resultant lower bounds of the different parameter measurement uncertainties ($\frac{\Delta\theta_{i}}{\theta_{i}}$). 

\begin{equation}
    F_{ij} = \int_{f_{low}}^{f_{high}} \frac{df}{S_{n}(f)} \langle \frac{\partial h}{\partial \theta_{i}}, \frac{\partial h}{\partial \theta_{j}} \rangle 
\end{equation}

\begin{equation}
        \frac{\Delta \theta_{i}}{\theta_{i}} = \sqrt{F_{ii}^{-1}}
\end{equation}

Current GW data analysis tools have mostly been developed to understand signals detected by detectors on Earth, and developments are being made to ease the LISA data analysis challenge~\citep{PhysRevD.103.083011, Hoy:2024ovd, Jan:2024zhr, PhysRevD.111.042009, Wu:2025zhc}. For Moon-based detectors, significant developments are yet to be made for a full Bayesian analysis, as the proposal is fairly recent. Hence, in this work, we focus on the results from the Fisher-matrix formalism. 
 
\section{Results} \label{sec:results}

In this Section, we focus on a GW170817-like BNS system and discuss the SNR, horizon distance, and early warning capabilities of different detector configurations. The true parameter values of such a source are listed in Table~\ref{tab:GW170817_params}.

\subsection{Early warning and Horizon distance}

A significant early warning alert of the BNS merger is necessary to prepare for observations across the entire EM spectrum. The time to merger/coalescence ($\tau_{c}$) is computed using Eq.4 from ~\cite{Chan:2018csa}:
\begin{equation}
\tau_{c} = \frac{5}{256}\frac{c^5}{G^{\frac{5}{5}}}\frac{(\pi f_{s})^{-\frac{8}{3}}}{\mathcal{M}^{\frac{5}{3}}}
\end{equation}
where $c$ is the speed of light, $G$ the gravitational constant, and $f_{s}$ the starting frequency considered for the GW signal. 

For a detector to claim a GW signal detection, a general SNR threshold of 8 needs to be attained. Figure~\ref{fig:EW} shows the SNR of different detectors as a function of time to merger (in hours). \texttt{Lunar-SEI} can claim a detection $\sim$3 years before merger.  {We choose an $f_{low}$ of 0.1 Hz for \texttt{Lunar-SEI} and \texttt{LILA-Horizon} as it is the lower bound of the mid-band GW frequency (0.1-10 Hz). A GW170817-like system is $\sim$7 years from merger at this frequency. \texttt{LILA-Horizon} also sees the signal on a similar timescale as \texttt{Lunar-SEI} but the SNR increases substantially as the system gets closer to merger}.  \texttt{Lunar-SUS} on the other hand attains an SNR of 8 at $\sim$40 days before merger providing more than 1 month alert. The signal reaches SNR=8 in \texttt{ET} around 15 hours before the merger. \texttt{CE40} can claim a detection only around 1 hour before the merger, and \texttt{LHO} only about 5 minutes before the merger. \\

\begin{figure}
    \centering
    \includegraphics[width=0.9\linewidth]{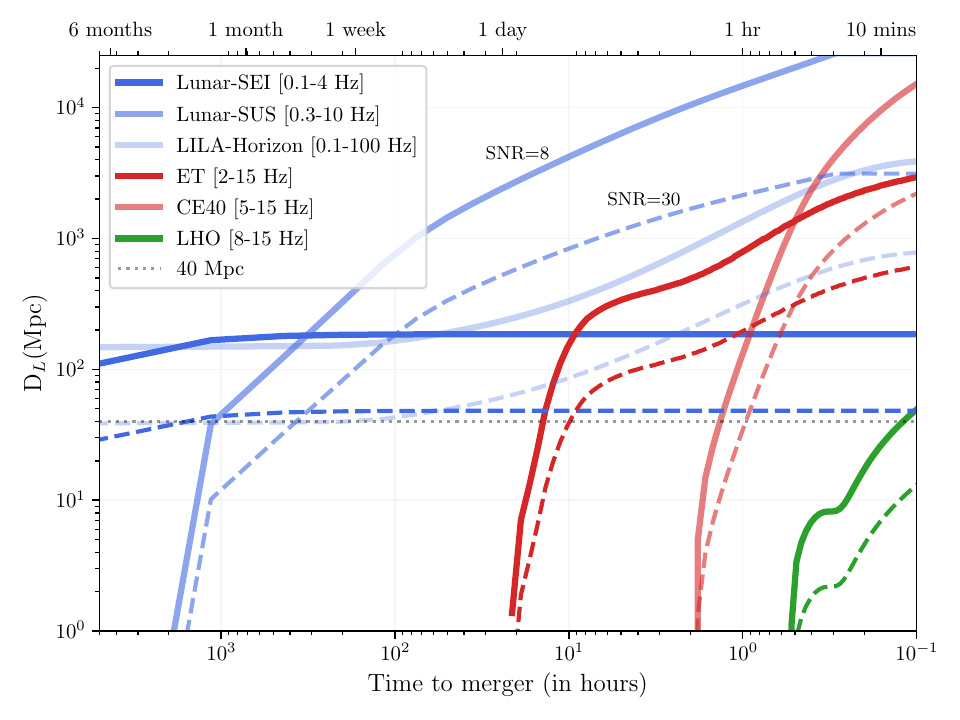}
    \caption{Horizon distance of a GW170817-like system as a function of time to merger in hours, for a threshold of SNR=8 (solid curve) and SNR=30 (dashed curve). The gray dotted line denotes the reported luminosity distance of GW170817 (40 Mpc).}
    \label{fig:Dl_tmerger}
\end{figure}

Assuming optimal sky location and orientation, we compute the horizon distance of a source with intrinsic parameters similar to GW170817 for various detector configurations at a detection threshold of SNR=8 and SNR=30, as shown in Figure~\ref{fig:Dl_tmerger}. At SNR=8, \texttt{Lunar-SEI} can detect sources a few years before merger but only till $\sim$200 Mpc, whereas \texttt{Lunar-SUS} can claim a detection $\sim$ 4 days before merger, up to a distance of 1 Gpc. {\texttt{LILA-Horizon} can see such a signal till 300 Mpc about 10 hours before merger, at SNR=8}. \texttt{ET} can give a $\sim$10 hour alert for the same horizon distance as observed by \texttt{Lunar-SEI}, but its reach increases as the system gets closer to merger. \texttt{CE40} on the other hand can detect GW170817-like systems around 1-2 hours before merger to $\sim$ 200 Mpc, and its reach increases significantly to $\sim$ 2 Gpc for a system that is 30 minutes from merger. When the SNR threshold is increased to 30, the horizon distance reduces by an order of magnitude for most detectors.

\begin{figure}
    \centering
 \includegraphics[width=0.8\linewidth]{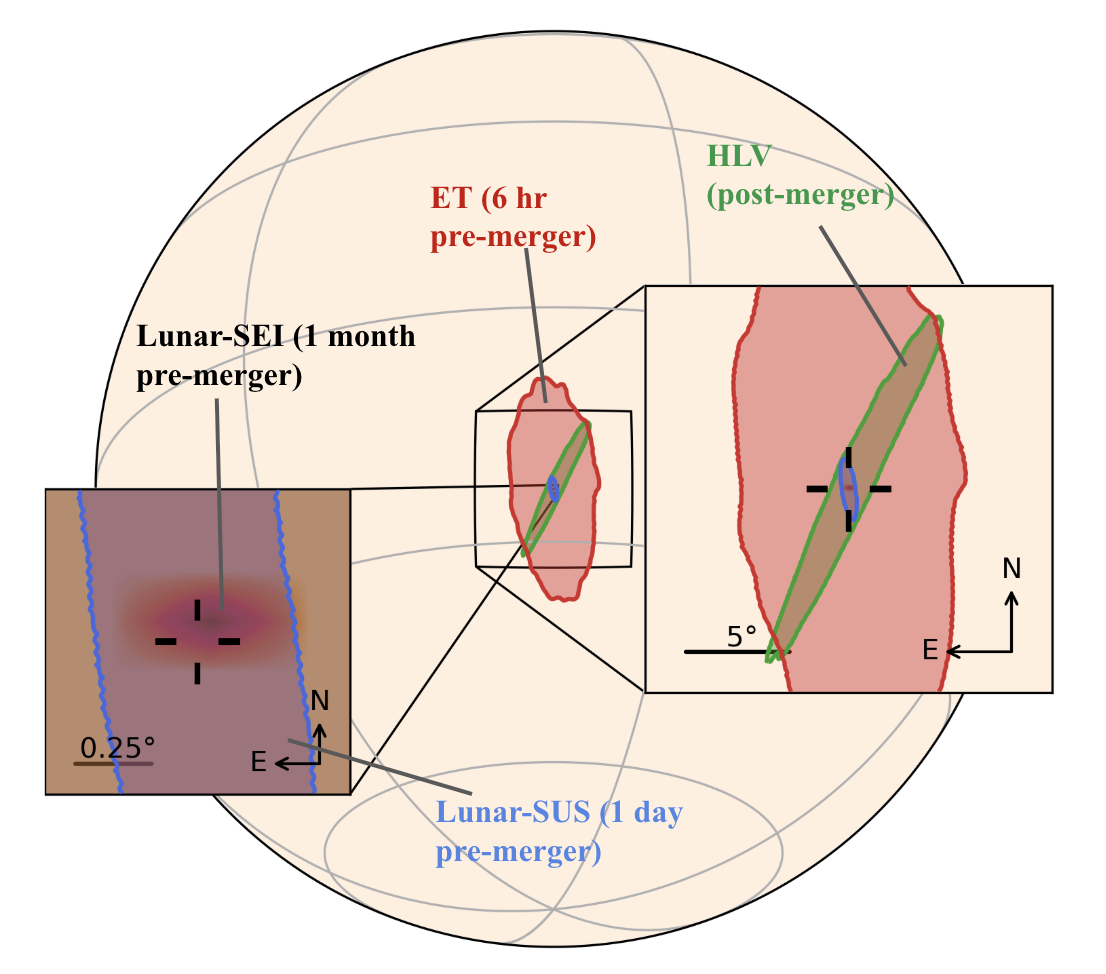}
    \caption{90$\%$ CI of the sky location errors ($\Delta\Omega$) for a GW170817-like system at luminosity distance of 40 Mpc, observed in \texttt{Lunar-SEI} at 1 month pre-merger (shaded region in the left inset near the marker), \texttt{Lunar-SUS} at 1 day pre-merger, \texttt{ET} at 6 hours pre-merger along with HLV sky localization of the real event (includes the complete observed inspiral signal).}
    \label{fig:LunarSUS_HLV}
\end{figure}

\subsection{Early warning and Sky localization}

For confident EM counterpart identification, an early warning alert should be accompanied by accurate source localization information. For this, we compute the sky localization errors of a GW170817-like system as it approaches merger, for various detector configurations. \texttt{Lunar-SEI} localizes a GW170817-like source to within $\sim$5 {arcmin$^{2}$} almost 1 month before merger. \texttt{Lunar-SUS} on the other hand localizes GW170817 to $\sim$95.6 deg$^{2}$ about 1 month before merger and $\sim$1.5 deg$^{2}$ about 1 day away from merger. Figure~\ref{fig:LunarSUS_HLV} shows the sky localization of a GW170817-like source for different Lunar detectors and \texttt{ET} along with the Hanford-Livingston-Virgo(HLV) skymap for GW170817~\citep{LIGOScientific:2017vwq}. The \texttt{HLV} skymap (28 deg$^2$) is from the real observations of GW170817 which was computed after the signal was out-of-band and not in the early warning phase. The \texttt{ET} 6-hour pre-merger skymap is $\sim$ 100 deg$^2$. We see that the Lunar detectors outperform any Earth-based detectors in this aspect.

\begin{figure}
    \centering
    \includegraphics[width=0.85\linewidth]{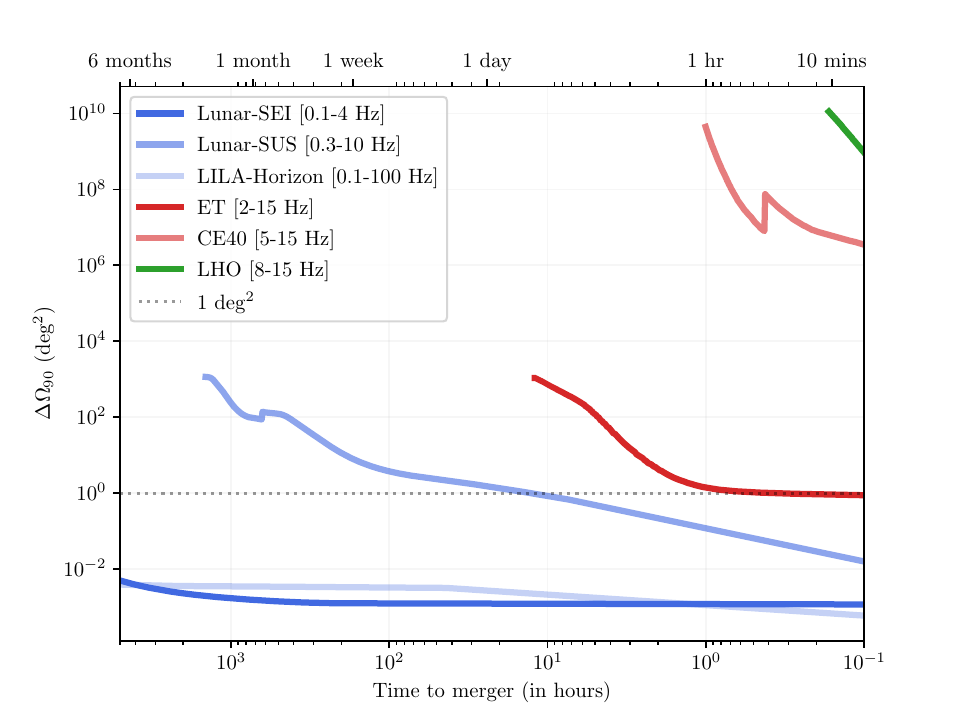}
    \caption{90$\%$ CI of the sky location errors ($\Delta\Omega$) of a GW170817-like system at a luminosity distance of 40 Mpc, as a function of time to merger in hours. Note this is the early warning alert sky localization.}
    \label{fig:Skyloc_tmerge}
\end{figure}

We also show how the sky localization improves for various detectors as the system gets closer to merger, in Fig.~\ref{fig:Skyloc_tmerge}. Due to the lower frequency sensitivity of \texttt{Lunar-SEI}, it localizes a GW170817-like source to under 0.01 deg$^2$ more than 6 months before merger, {similar to \texttt{LILA-Horizon}}.  \texttt{Lunar-SUS} starts to localize the source to within 10 deg$^2$ more than 1 week from merger, and to less than 1 deg$^2$ when it is less than 1 day away from merger.  

The sky localization accuracy obtained by the Lunar detectors 1 day before merger falls well within the field of view (FOV) of EM follow-up telescopes. Some of the telescopes that observed GW170817 are listed as follows with their FOV in brackets: Swift (0.15 deg$^{2}$)~\cite{10.1117/12.409158}, Chandra (0.15 deg$^{2}$)~\cite{Weisskopf:1999hm}, Hubble Space Telescope (0.0031 deg$^{2}$), Very Large Array (0.0625 deg$^{2}$). We also list some of the new telescopes that were launched after GW170817 as well as upcoming telescopes: James Webb Space Telescope (0.0027 deg$^{2}$), Rubin Observatory (9.6 deg$^{2}$),  Roman (0.28 deg$^{2}$), and Athena (0.35 deg$^{2}$).

\begin{table}[t!]
\centering
\renewcommand{\arraystretch}{1.3}
\setlength{\tabcolsep}{6pt}
\begin{tabular}{lccc|ccc|ccc}
\toprule
 & \multicolumn{3}{c}{\textbf{1 day}} & \multicolumn{3}{c}{\textbf{1 hour}} & \multicolumn{3}{c}{\textbf{10 minutes}} \\
\cmidrule(lr){2-4}\cmidrule(lr){5-7}\cmidrule(lr){8-10}
\textbf{Network} & $<100$ & $<10$ & $<1$ & $<100$ & $<10$ & $<1$ & $<100$ & $<10$ & $<1$ \\
 & (deg$^2$) & (deg$^2$) & (deg$^2$) & (deg$^2$) & (deg$^2$) & (deg$^2$) & (deg$^2$) & (deg$^2$) & (deg$^2$) \\
\midrule
\texttt{Lunar-SEI}  & 20   & 20   & 20   & 20    & 20    & 20   & 20    & 20    & 20 \\
\texttt{Lunar-SUS}  & 394  & 98   & 22   & 8524  & 4459  & 825  & 11300 & 5957  & 3678 \\
{\texttt{LILA-Horizon}} & {27} & {27} & {27} & {699} & {679} & {487} & {4857} & {4601} & {1122} \\ 
\texttt{ET+CE40+L}    & 0    & 0    & 0    & 48    & 10    & 1    & 948   & 92    & 19 \\
\texttt{HLA}        & 0    & 0    & 0    & 0     & 0     & 0    & 0     & 0     & 0 \\
\bottomrule
\end{tabular}
\caption{\textit{Annual detection rates} of BNS mergers for different detector networks. 
Entries show the number of mergers expected per year that can be detected with early-warning times of 1 day, 1 hour, or 10 minutes before merger. 
Within each block, the columns correspond to the number of events localized to within sky areas of $<100$, $<10$, or $<1$ deg$^2$ (90\% credible region). 
For example, the row for Lunar-SUS indicates that $\sim$100 events per year could be localized to better than $10$ deg$^2$ with 1 day of advance warning. 
}
\label{tab:EW_skyloc}
\end{table}

\subsection{Detection rates}

Identifying an EM counterpart requires a significant early warning alert, accompanied by an accurate understanding of the source's sky localization. To obtain detection rates for Lunar configurations, we follow the methods used in Pandey et al. ~\citep{Pandey:2024mlo}, and distribute a population of neutron star binaries across the Universe with the component masses $m_{1}$, $m_{2}$ $\in$ U[1,2.2]$M_{\odot}$, component spin magnitudes $a_{1},a_{2}$  $\in$ U[0,0.1]. The orientation of the binary is chosen such that cos($\iota$) $\in$ U[-1,1] and $\psi$ $\in$ U[0,2$\pi$]. The sky location of the binary is chosen such that RA $\in$ U[0,2$\pi$] and cos(DEC) $\in$ U[-1,1]. The redshift distribution is populated till z=0.5, and follows the Madau-Dickinson star formation rate~\citep{Madau:2014bja} and a local merger rate of $\dot{n}$(0) = 320 Gpc$^{-3}$yr$^{-1}$ is assumed.  The luminosity distance is computed from the redshift assuming the Planck18 cosmology in \texttt{Astropy}. $\phi_{c}$ is fixed at 0. We then use \texttt{GWFish} to compute SNR and sky localization errors for the populated sources. 


In Table~\ref{tab:EW_skyloc}, we show annual detection rates for select detector configurations with different early warning alert times. We see that the Lunar detectors alone can detect tens of mergers with a 1-day early warning alert. \texttt{Lunar-SUS} sees about 100 binaries that have sky localization area under 10 deg$^{2}$. \texttt{Lunar-SEI} on the other hand, always sees the same number of $\sim$20 mergers that are well localized due to its distance reach. {At 1 day before merger, \texttt{LILA-Horizon} sees a comparable number of sources to \texttt{Lunar-SEI}, but as the early warning time gets closer to merger the detection rate increases significantly.} \texttt{HLA} does not see any mergers with even a 10-minute early warning alert. 3G detectors (\texttt{ET+CE40+L}) also do not provide a 1 day early warning alert with a good sky localization for any BNS mergers. But they do see a few mergers that are localized with less than 1 deg$^{2}$ accuracy about 1 hour before merger. This number increases as the early warning alert time reduces to 10 minutes. Hence, Lunar detectors uniquely deliver \textit{hundreds of well-localized events} annually; Earth-only detectors cannot.

\subsection{Parameter constraints and Multiband parameter estimation}

In this section, we list the constraints on the errors of select BNS parameters of a GW170817-like source for different detector configurations using the Fisher formalism with \texttt{IMRPhenomD$\_$NRTidalv2}~\citep{Dietrich:2019kaq} waveform model. 

\begin{table}[t!]
\centering
\renewcommand{\arraystretch}{2} 
\setlength{\tabcolsep}{5pt}
\begin{tabular}{lcccccccc }
\hline
\textbf{Network} & SNR & $\sigma_{\mathcal{M}}$ & $\sigma_{q}$ & $\sigma_{\tilde{\Lambda}}$ & 
$\sigma_{a_{1}}$ & $\sigma_{D_{L}}$ & $\Delta\Omega$ & $\iota$\\
& & ($M_{\odot}$) &  & & & (Mpc) & (deg$^2$)  & (deg)\\
\hline
\texttt{Lunar-SEI} & 28  & 2$\times10^{-6}$  & 2$\times10^{-1}$  & 2$\times10^{5}$  & 7$\times10^{-3}$  & 4$\times10^{1}$  & 8$\times10^{-3}$ & 70\\

\texttt{Lunar-SUS} & 1458  & 5$\times10^{-7}$  &  2$\times10^{-2}$  & 2$\times10^{5}$  & 6$\times10^{-4}$  & 1  & 2$\times10^{-2}$ & 2\\
\hline
\texttt{Lunar-SEI+HLA} & 151  & 6$\times10^{-8}$  & 7$\times10^{-3}$  & 7$\times10^{1}$  & 4$\times10^{-1}$  & 6  & 6$\times10^{-4}$ & 10\\

\texttt{Lunar-SUS+HLA} & 1466  & 5$\times10^{-8}$  & $10^{-3}$  & 3$\times10^{1}$  & 4$\times10^{-5}$  & 5$\times10^{-1}$  & 5$\times10^{-7}$ & 1\\

\texttt{Lunar-SEI+SUS+HLA} & 1466 & 10$^{-8}$ & 6$\times10^{-4}$ & 2$\times10^{1}$ & 10$^{-5}$ & 5$\times10^{-1}$ & 4$\times10^{-7}$ & 1 \\
\hline
\texttt{ET+CE40+CE20(Aus)+A}  & 2246  & 7$\times10^{-7}$  & $10^{-2}$  & 6  & $10^{-1}$  & 1  &  3$\times10^{-3}$ &  2\\

\texttt{ET+HLA} & 662  & $10^{-6}$  & 2$\times10^{-2}$  & 3$\times10^{1}$  &  4$\times10^{-1}$  & 1 & 2$\times10^{-2}$ & 3\\

\hline
\end{tabular}

\caption{90\% credible levels for errors on various parameters for a GW170817-like source. fmax=2kHz for all detector configurations except \texttt{Lunar-SUS}=10Hz and \texttt{Lunar-SEI}=4Hz. $\sigma_{\mathcal{M}}$ is in units of $M_{\odot}$, $\sigma_{D_{L}}$ is in Mpc, $\Delta\Omega$ is in $deg^{2}$, inclination angle($\iota$) is in $deg$, and rest of the errors are dimensionless.} 
\label{tab:detconfigerrorsv6}
\end{table}

In Table ~\ref{tab:detconfigerrorsv6}, when considering the Lunar-only detectors, we see that \texttt{Lunar-SUS} outperforms \texttt{Lunar-SEI} in terms of SNR and an order of magnitude better constraint in mass and spin parameters, but \texttt{Lunar-SEI} provides an order of magnitude better sky localization error. We also see that \texttt{Lunar-SUS} can constrain the luminosity distance and inclination angle errors at least an order of magnitude better than \texttt{Lunar-SEI}. Comparing with the Earth-based detectors and a network of Lunar+Earth detectors, we see that the 2G+3G detector configurations \texttt{ET+CE40+CE20(Aus)+A} as expected, detects the signal with a much larger SNR than the current 2G detector configuration of \texttt{HLA}, with two orders of improvement on chirp-mass errors. Performing this analysis for a joint Lunar+HLA configuration increases the SNR considerably, with the error constraints being smaller by multiple orders of magnitude, for both \texttt{Lunar-SEI} and \texttt{Lunar-SUS}. 
 
\begin{figure}[t!]
    \centering
 \includegraphics[width=0.7\linewidth]{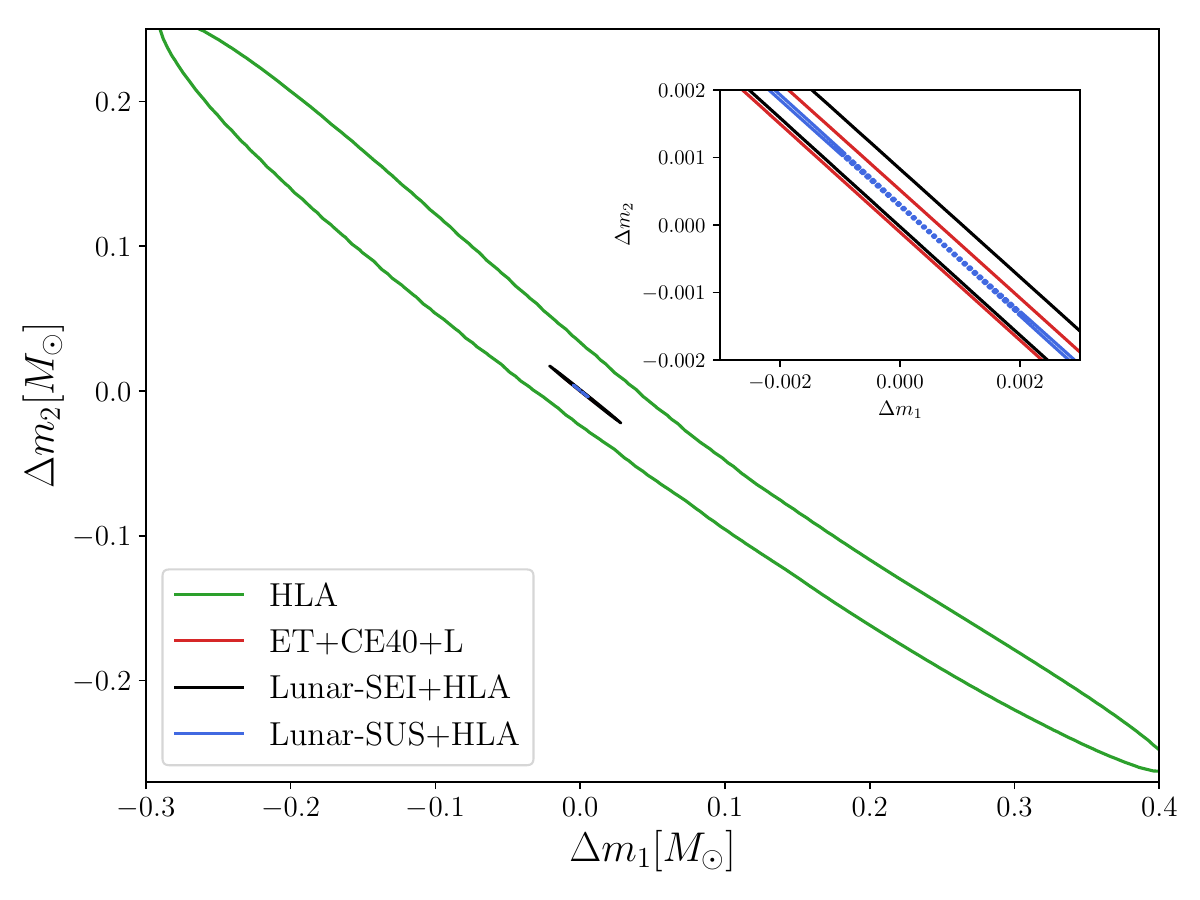}
    \caption{90$\%$ CI posteriors for an Earth-only 2G analysis (\texttt{HLA}), Earth-only 2G and 3G detectors (\texttt{ET+CE40+L}), and Multiband (\texttt{Lunar-SEI+HLA}, \texttt{Lunar-SUS+HLA}) analysis. The blue star denotes the true value of the parameters. We plot the difference of the individual mass posteriors from their true value, for example $\Delta m_{1}=m_{1} - \bar{m_{1}}$.}
    \label{fig:Multiband}
\end{figure}

We now discuss constraints on masses of individual neutron stars for the Lunar+Earth-2G networks and for the Earth-based detectors, we consider two networks- one with 2G-only detectors (\texttt{HLA}) and second which also includes 3G detectors (\texttt{ET+CE40+L}). Figure~\ref{fig:Multiband} shows the parameter constraints of a  GW170817-like event in \texttt{HLA}, \texttt{ET+CE40+L}, \texttt{Lunar-SEI+HLA} and \texttt{Lunar-SUS+HLA} posteriors for the masses. We see that the \texttt{Lunar-SUS+HLA} configuration analysis provides stronger constraints in the mass space, even better than 3G Earth-based detectors. 

A single Lunar detector combined with current 2G detectors provide similar and sometimes better constraints compared to a network of 3G detectors on Earth in the parameter space of neutron stars. But it is important to re-iterate here that finding EM counterparts of such mergers is equally important in confirming that they are indeed neutron stars, and the Lunar detectors surpass any Earth-based 3G detector in that. Similar multiband analyses have been demonstrated for high-mass binary black hole systems that could potentially be observed by both the space-based GW detector LISA (Laser Interferometer Space Antenna) and the terrestrial LVK detectors~\citep{PhysRevLett.117.051102, Sesana:2017vsj, Grimm:2020ivq, Ranjan:2024wui, Ruiz-Rocha:2024xjt, Wu:2025zhc}.


From Table ~\ref{tab:detconfigerrorsv8} we see that overall, multi-band Earth-Moon detector networks provide up to orders-of-magnitude improvement in both equation-of-state and Hubble constant constraints compared to Earth-only 3G facilities.

\begin{table}[h!]
\centering
\renewcommand{\arraystretch}{1.5} 
\begin{tabular}{lccc|ccc }
\hline
 & \multicolumn{3}{c}{\textbf{Equation of State}} & \multicolumn{3}{c}{\textbf{Multimessenger Follow-up }} \\
\textbf{Network} & $\sigma_{\mathcal{M}}$ ($M_{\odot}$) & $\sigma_{\tilde{\Lambda}}$ & 
$\sigma_{a_{1}}$ & $\sigma_{D_{L}}$ (Mpc)& $\Delta\Omega$ (deg$^2$) & $\iota$ (deg)\\
\cmidrule(lr){1-4} \cmidrule(lr){5-7}
\texttt{Lunar-SEI} & $10^{-6}$  & $10^{5}$  & $10^{-3}$  & $10^{1}$  & $10^{-3}$ & $10^{2}$\\

\texttt{Lunar-SUS} & $10^{-7}$  &  $10^{5}$  & $10^{-4}$  & 1  & $10^{-2}$ & 1\\
\hline
\texttt{Lunar-SEI+HLA} & $10^{-8}$  & $10^{1}$  & $10^{-1}$  & 6  & $10^{-4}$ & $10^{1}$\\

\texttt{Lunar-SUS+HLA} & $10^{-8}$  & $10^{1}$  & $10^{-5}$  & $10^{-1}$  & $10^{-7}$ & 1\\

\texttt{Lunar-SEI+Lunar-SUS+HLA} & 10$^{-8}$ & $10^{1}$ & 10$^{-5}$ & $10^{-1}$ & $10^{-7}$ & 1 \\

\hline
\texttt{HLA} & $10^{-5}$ & 50 & 2 & 10 & $10^{-1}$ & $10^{1}$ \\


\texttt{CE40+LA} & $10^{-6}$ & 7 & $10^{-1}$ & 8 & $10^{-1}$ & $10^{1}$\\

\texttt{ET+HLA} & $10^{-6}$  & $10^{1}$  &  $10^{-1}$  & 1 & $10^{-2}$ & 1\\
\hline
\texttt{ET+CE40+L} & $10^{-7}$ & 6 & 10$^{-1}$ & 1 & 10$^{-2}$ & 1\\

\texttt{ET+CE40+A} & $10^{-7}$ & 6 & 10$^{-1}$ & 1 & $10^{-3}$ & 1\\

\texttt{ET+CE40+LA} & $10^{-7}$ & 6 & 10$^{-1}$ & $10^{-1}$ & $10^{-3}$ & 1\\


\texttt{ET+CE40+CE20(Aus)+LA} & $10^{-7}$ & 6 & 10$^{-1}$ & $10^{-1}$ & $10^{-3}$ & 1\\


\texttt{ET+CE20(Aus)+HLA} & 10$^{-6}$ & 13 & $10^{-1}$ & 1 & $10^{-3}$ & 1 \\


\texttt{CE40+CE20(Aus)+LA} & $10^{-6}$ & 7 & $10^{-1}$ & 2 & $10^{-3}$ & 1 \\

\texttt{ET+CE40+CE20(Aus)+A} & $10^{-7}$  & 6  & $10^{-1}$  & 1  &  $10^{-3}$ & 1\\

\hline\hline
\end{tabular}

\caption{Comparison between detector networks for two key BNS science cases: 
(i) constraining the dense-matter equation of state (left block: chirp mass $\sigma_{\mathcal{M}}$, tidal deformability $\sigma_{\Lambda}$, spin $\sigma_{a_{1}}$), and 
(ii) measuring the Hubble constant through standard sirens (right block: luminosity distance $\sigma_{D_{L}}$, sky-localization area $\Delta\Omega$, inclination angle $\iota$). 
Values correspond to 90\% credible levels for a GW170817-like source, computed using Fisher-matrix parameter estimation. 
For all networks, $f_{\rm max}=2$ kHz except for \texttt{Lunar-SUS}=10 Hz and \texttt{Lunar-SEI}=4 Hz. 
$\sigma_{\mathcal{M}}$ is in units of $M_{\odot}$, $\sigma_{D_{L}}$ in Mpc, $\Delta\Omega$ in deg$^{2}$, inclination angle in degrees, and other quantities are dimensionless.}
\label{tab:detconfigerrorsv8}
\end{table}

\subsection{Equation of state}

{ In Tables 3 and 4, we see that while multi-band Earth-Moon detector networks achieve tidal deformability constraints ($\tilde{\Lambda}$) comparable to Earth-only 3G networks, they deliver a transformative advantage through 1-2 orders of magnitude improved constraints on chirp-mass and mass ratios—which are most tightly coupled to tidal effects in the frequency-domain phase evolution. For a GW170817-like event, the \texttt{Lunar-SUS+HLA} and \texttt{Lunar-SEI+HLA} configurations achieve chirp mass precision at the ~10$^{-8}$ M$_{\odot}$ level, breaking mass-tidal degeneracies and enabling significantly tighter posterior constraints on $\tilde{\Lambda}$ in realistic Bayesian analyses.}

\section{Discussion} \label{sec:discussion}

In summary, the integration of gravitational-wave observations from Earth- and Moon-based detectors marks a transformative step for multimessenger astronomy. This work provides a thorough quantitative demonstration of how a lunar gravitational-wave network expands the discovery space for BNSs. Lunar observatories can access the sub-hertz band unattainable on Earth, enabling observations of BNS mergers months to years before coalescence, compared to the hours-to-minutes horizon of even third-generation terrestrial facilities. This early access translates directly into superior sky localization: Lunar-SEI achieves localization at the level of $\sim$5 arcmin$^{2}$ one month before merger, while multi-band Earth–Moon networks reduce the localization to $\sim$6 arcsec$^{2}$, comparable to the field of view of the James Webb Space Telescope at high zoom.

These gains are accompanied by dramatic improvements in parameter estimation. For GW170817-like systems, multi-band networks measure the chirp mass to parts per billion, the mass ratio to $\sim$0.1$\%$ precision, and neutron star spins to better than 0.001$\%$. Luminosity distance uncertainties are reduced to the $\sim$1$\%$ level, enabling standard-siren determinations of the Hubble constant with unprecedented accuracy. Annual detection rates also increase substantially: lunar facilities alone yield tens to hundreds of well-localized mergers per year, enabling a new scale of coordinated EM and neutrino follow-up.

Our results demonstrate that lunar gravitational-wave observatories would revolutionize multi-messenger astrophysics with BNSs, transforming them into precision probes of nuclear physics and cosmology. In doing so, they define a unique discovery landscape that naturally aligns with NASA’s Artemis exploration program and establishes the scientific case for LILA as a cornerstone of the next era of gravitational-wave astronomy.

\ack{A. B. Y. and K. J.’s work was supported in part by the Lunar Labs Initiative at Vanderbilt
University, which is funded by grants from the REAM Foundation and the Vanderbilt Scaling Grant from the Office of the Vice Provost for Research and Innovation. This work makes use of the software \texttt{GWFish}~\citep{DupletsaHarms2023} https://github.com/AnjaliYelikar/GWFish} 


\bibliography{paper}

\end{document}